\documentclass[preprint]{aastex}
\usepackage{amssymb}
\usepackage{amsmath}
\usepackage{graphicx}

\newcommand{\txd}{{\text{d}}}

\newcommand{\txV}{{\text{V}}}

\begin{document}

\title{Efficient radiative transfer modelling with SKIRT}

\author{Maarten Baes}
\affil{European Southern Observatory, La Casilla 19001, Santiago 19, Chile}
\affil{Sterrenkundig Observatorium, Universiteit Gent, Krijgslaan 281 S9, B-9000 Gent, Belgium} 
\email{mbaes@eso.org, maarten.baes@ugent.be}
\and
\author{Herwig Dejonghe}
\affil{Sterrenkundig Observatorium, Universiteit Gent, Krijgslaan 281 S9, B-9000 Gent, Belgium}
\email{herwig.dejonghe@ugent.be}
\and
\author{Jonathan I.\ Davies}
\affil{Cardiff University, 5 The Parade, Cardiff CF24\,3YB, Wales}
\email{jonathan.davies@astro.cf.ac.uk}

\begin{abstract}
We present SKIRT, a three-dimensional Monte Carlo radiative transfer
code developed to study dusty galaxies. We discuss SKIRT's most
important characteristics and present a number of applications. In
particular, we focus on the kinematical aspect of SKIRT. We
demonstrate that dust attenuation mimics the presence of dark matter
around elliptical galaxies and that it severely affects the rotation
curves of edge-on galaxies.
\end{abstract}


\section{Introduction}

In many astronomical situations dust plays a crucial role in our
interpretation of observations. In these situations the inferences
made from data may require accurate modelling of dust absorption,
scattering and/or emission mechanisms. This modelling is not
straightforward and is often neglected in the hope that it is not
important. Only during the past few years, radiative transfer codes
have become powerful enough to handle realistic two- or
three-dimensional geometries. Many different approaches exist to
handle the radiative transfer problem, and various authors have
adopted different techniques to study the effect of dust on the
photometry and SEDs of galaxies (e.g.\ \cite{1988ApJ...333..673B},
\cite{1992ApJ...393..611W}, \cite{1994ApJ...432..114B},
\cite{1994A&A...282..407K}, \cite{1996ApJ...461..155W},
\cite{1996ApJ...465..127B}, \cite{1998ApJ...509..103S},
\cite{2000A&A...362..138P}, \cite{2000A&A...359...65B},
\cite{2001MNRAS.326..722B}, \cite{2001MNRAS.326..733B},
\cite{2001ApJ...551..277M}, \cite{2003A&A...407..159G}). In this
contribution, we present a three-dimensional Monte Carlo radiative
transfer code SKIRT (Stellar Kinematics Including Radiative
Transfer). The code was initially developed to study the observed
kinematics of dusty galaxies, but it has now turned into a very
flexible and general tool to study various radiative transfer problems
in dusty systems.

\section{Basic characteristics of SKIRT}

The basic characteristics of Monte Carlo radiative transfer are
described in e.g.\ \cite{CE59}, \cite{1970A&A.....9...53M},
\cite{1977ApJS...35....1W}, \cite{1994A&A...284..187F},
\cite{1996ApJ...465..127B}, \cite{2001ApJ...551..269G} and
\cite{2003A&A...399..703N}. The key ingredient in Monte Carlo
radiative transfer is that the radiation field is treated as a flow of
a finite number of monochromatic photon packages. At each moment in
the simulation, a photon package is characterized by a luminosity, a
frequency, a position and a propagation direction. A Monte Carlo
radiative transfer simulation consists of consecutively following the
individual path of each single photon package through the interstellar
medium.  The trajectory of the photon package is determined by various
events such as emission, absorption and scattering events. Each event
is generated randomly from the appropriate probability
distribution. At the end of the simulation, knowledge of the
individual path of each photon package is used to completely determine
the radiation field at all positions, in all directions and at all
frequencies.

Monte Carlo codes are very flexible, they allow arbitrary geometries
for stars and dust and they are easy to interpret and to
implement. The main disadvantage is also well-known: Monte Carlo codes
are relatively slow and numerically demanding compared to other
techniques. Our experience is, however, that this disadvantage is
negligible compared to the benefits of the method, especially as Monte
Carlo simulations can nowadays be optimized substantially by the
inclusion of deterministic elements. Many of these elements are
implemented into the SKIRT code, such as the forced first scattering
principle \cite{CE59}, the peel-off technique
\cite{1984ApJ...278..186Y} and continuous absorption
\cite{1999A&A...344..282L}.  A novel and very useful feature is the
frequency distribution adjustment principle introduced by Bjorkman \&
Wood \cite{2001ApJ...554..615B}. This powerful technique ensures that
at each moment during a Monte Carlo simulation, the dust temperature
and the local radiation field are always in agreement. As a result,
Monte Carlo simulations can self-consistently calculate the dust
temperature distribution and the far-infrared emission of galaxies
without the costly iteration that is a necessary ingredient in most
radiative transfer schemes. For a critical discussion, we also refer
to \cite{B+04}. Apart from these optimization techniques borrowed and
adapted from other Monte Carlo modellers, we have included two novel
mechanisms in the SKIRT code: the use of partly polychromatic photon
packages and the inclusion of kinematic information.

\section{Polychromatic photon packages}

The photon packages used in most Monte Carlo simulations are
monochromatic, i.e.\ they consist of a unspecified number of
individual photons with the same frequency. This has a good reason:
many of the probability functions which determine the trajectory of a
photon package are frequency-dependent. For example, the probability
distribution of the covered path length before an interaction is an
exponential distribution in optical depth space, and therefore a
frequency-dependent distribution in physical path length. Consequences
of this monochromatism are a low signal-to-noise of the radiation
field in frequency regions where the stars and/or dust are only poor
emitters and a noisy dust temperature distribution. SKIRT works with
polychromatic photon packages, i.e.\ packages characterized by a
continuous frequency distribution rather than by a single
frequency. The photon packages are only partly polychromatic: they are
polychromatic when they are emitted (by either the stars or the dust
grains), but loose their polychromatism during a scattering event. In
combination with the peel-off technique and the treatment of
absorption as a continuous process, this mechanism assures that every
single photon package contributes at least once to the observed
radiation field at every single wavelength, and that the absorption
rate is determined with much greater precision. As a result, much less
photon packages (and thus CPU time) are necessary to obtain high
signal-to-noise results. 

\section{The inclusion of kinematic information}

\begin{figure}[t!]
\centering
\includegraphics[bb=80 210 504 703,clip,width=0.5\textwidth]{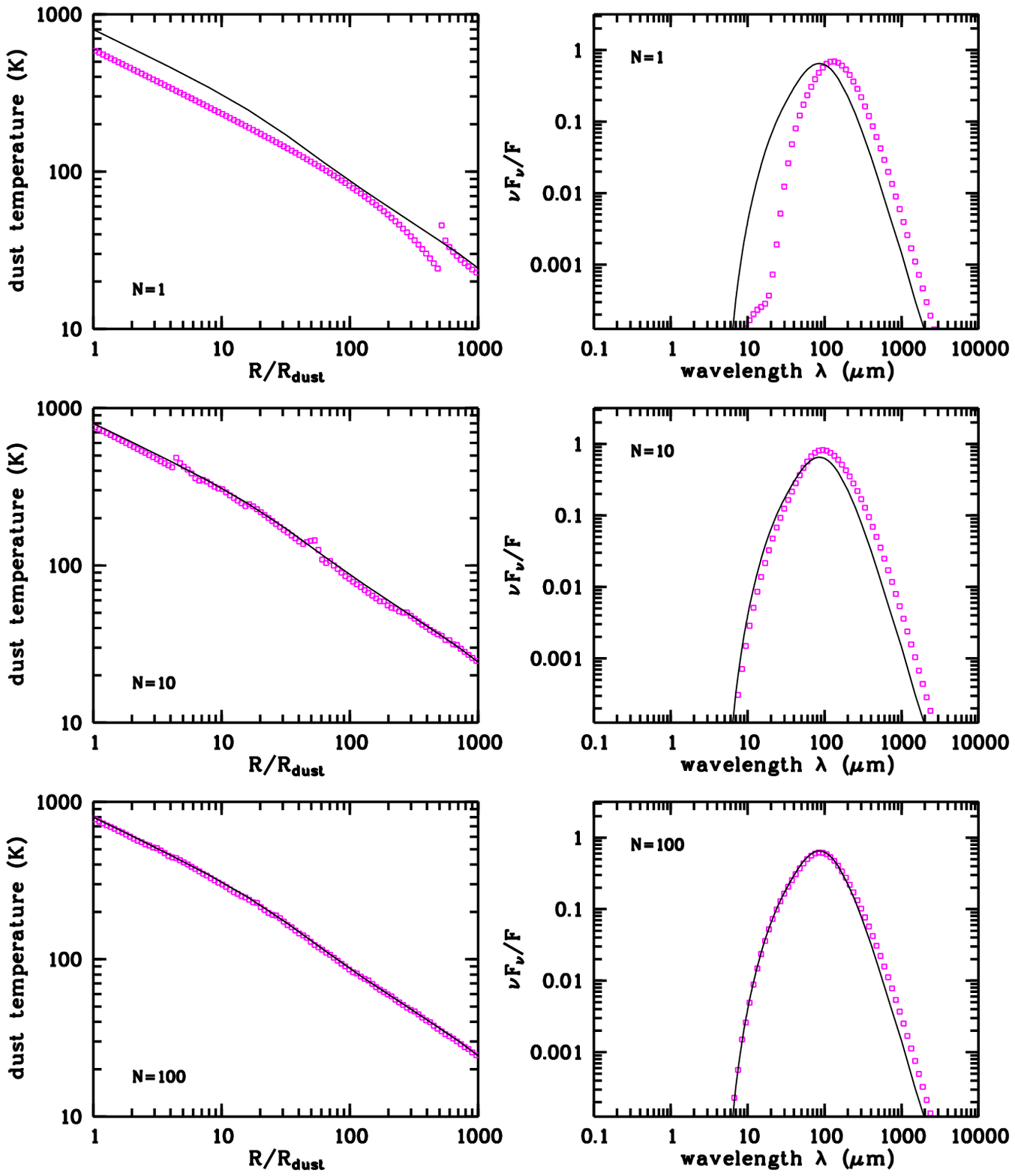}%
\includegraphics[bb=80 210 504 703,clip,width=0.5\textwidth]{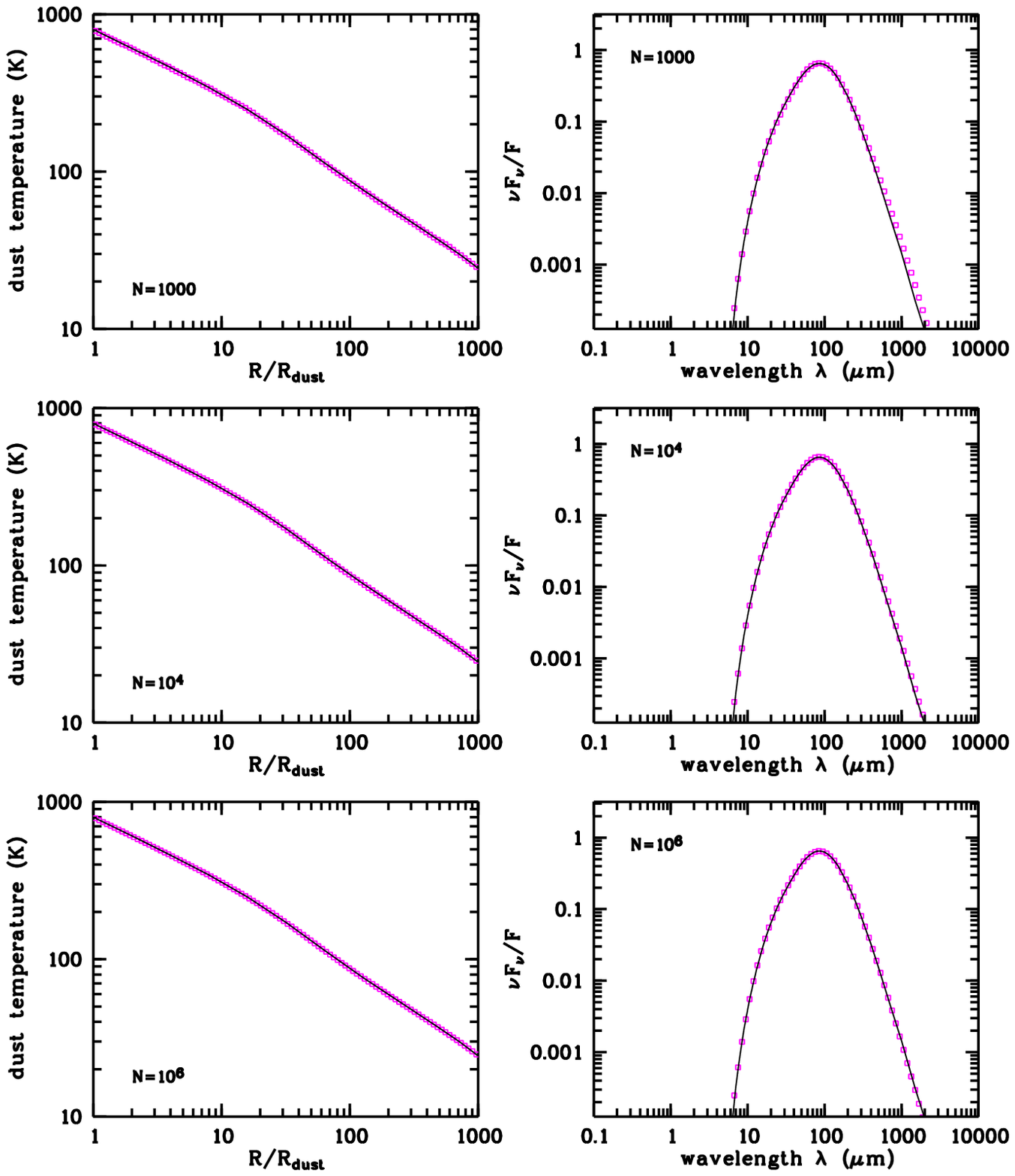}
\caption{Demonstration of the efficiency of the SKIRT code to perform
LTE radiative transfer calculations. Shown are the temperature profile
(left) and the emerging spectrum (right) as obtained with the SKIRT
code for the $p=0$ and $\tau_\txV=100$ benchmark model from Ivezi\'c
et al.~\cite{1997MNRAS.291..121I}. In each panel, the grey dots are
the SKIRT results and the solid black lines represent the benchmark
solution. The six simulations presented correspond to the same model,
but the number of photon packages in the simulation ranges from 1 (top
left) to 10$^6$ (bottom right). Even with very few photon packages,
SKIRT is able to provide good results, thanks to the polychromatic
nature of the photon packages.}
\label{f1.eps}
\end{figure}

Kinematical information on external galaxies is obtained from
extracting line-of-sight velocity information from spectra, through
measuring the Doppler shifts of either absorption lines (stars) or
emission lines (gas). Often these spectra are taken in the optical
regime and therefore they are subject to dust attenuation. The effect
of dust attenuation on the observed kinematics is still largely
unexplored. Most studies focused on the effect of dust absorption on
the rotation curves of spiral galaxies (\cite{1990MNRAS.245..350D},
\cite{1992ApJ...400L..21B}, \cite{2001ApJ...548..150M},
\cite{2003AstL...29..437Z}, \cite{2004AJ....128..115V}). In order to
investigate in a more general way how the observed kinematics of
galaxies are affected by dust attenuation (including scattering), we
have included the possibility to include kinematical information in
the SKIRT code. This is accomplished by assigning to each photon
package an additional characteristic: the line-of-sight velocity it
carries with it. This characteristic is initialized according to the
emitting star's velocity, and is updated during every scattering
event. Note that this is not a trivial process because the individual
velocities of both the stars and the scattering dust grains need to be
taken into account. In order to increase the efficiency of the Monte
Carlo simulation, we implemented a similar technique as the
polychromatic photon packages: instead of assigning a single
line-of-sight velocity to each photon package, photon packages are
characterized by a line-of-sight velocity distribution. The
combination of this characteristic with the peel-off and forced
scattering techniques guarantees that every single photon contributes
several times to the observed kinematics at all possible line-of-sight
velocities, such that reliable results can be obtained in a reasonable
CPU time. For technical details on the implementation we refer to
\cite{2002MNRAS.335..441B} and \cite{2003MNRAS.343.1081B}.

\section{Benchmark results}

\begin{figure}[t!]
\centering
\includegraphics[bb=127 88 502 863,clip,angle=-90,width=\textwidth]{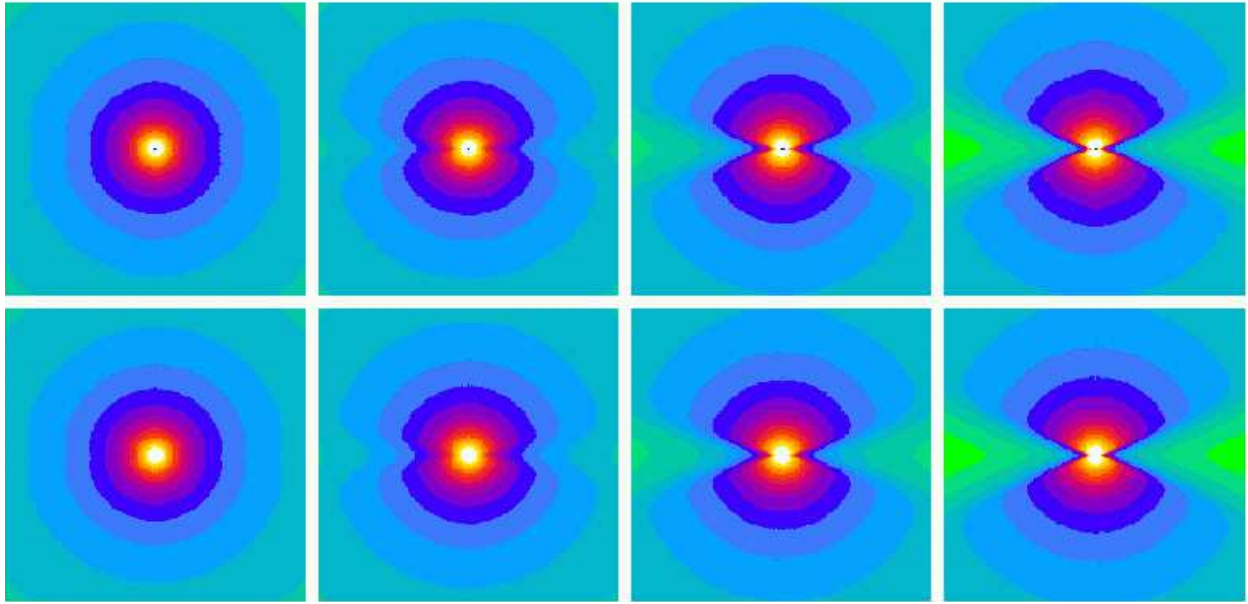}
\caption{SKIRT benchmark tests for the 2D axisymmetric benchmark
models from Pascucci et al.~\cite{2004A&A...417..793P}. The top panels
shows the temperature maps of the central $100\times100$~AU for the
models corresponding to $\tau_\txV=0.1$, 1, 10 and 100 (from left to
right), as obtained with RADICAL. The bottom panels show the
corresponding temperature maps obtained with SKIRT with $10^6$ photon
packages. The agreement is very satisfactory and lies within the
uncertainties quoted in \cite{2004A&A...417..793P}.}
\label{f2.eps}
\end{figure}

The only way to test the correct behaviour of a radiative transfer
code is by comparing its results with those of other codes on a single
well-defined benchmark model. Ivezi\'c et
al.~\cite{1997MNRAS.291..121I} provide the first benchmark solution
for a continuum LTE radiative transfer model. Their model consists of
a single star surrounded by a spherical dusty envelope with optical
depths ranging from $\tau_\txV=1$ to $\tau_\txV=1000$.  We have
calculated the dust temperature distribution and the emerging spectrum
of this benchmark model using the SKIRT code in both spherical and
axisymmetric geometries. The agreement with the benchmark results,
both for the temperature distribution and the spectral energy
distribution, is excellent, for all optical depths. In
figure~{\ref{f1.eps}} we compare one benchmark model with the results
of SKIRT simulation with different numbers of photon packages. Thanks
to the polychromatic nature of the photon packages, decent results are
already obtained with very few photon packages. Recently, Pascucci et
al.~\cite{2004A&A...417..793P} provided a more challenging
axisymmetric benchmark model, consisting of a star surrounded by an
dusty accretion disk with strong density gradients and optical depths
ranging from $\tau_\txV=0.1$ to $\tau_\txV=100$. They calculated
temperature maps and SEDs for these models using five different codes,
and found agreement to within some 10 percent. In
figure~{\ref{f2.eps}} we compare the temperature maps obtained with
SKIRT to those obtained with RADICAL, one of the radiative transfer
codes used in the benchmark calculations. The agreement between the
results is obvious, both for small and large optical depths. The SKIRT
simulations to obtain these results took only some 3 hours on a single
processor PC.

\section{Application~1: Edge-on spiral galaxy SEDs}

\begin{figure}[t!]
\includegraphics[bb=81 26 598 766,clip,height=0.8\textheight]{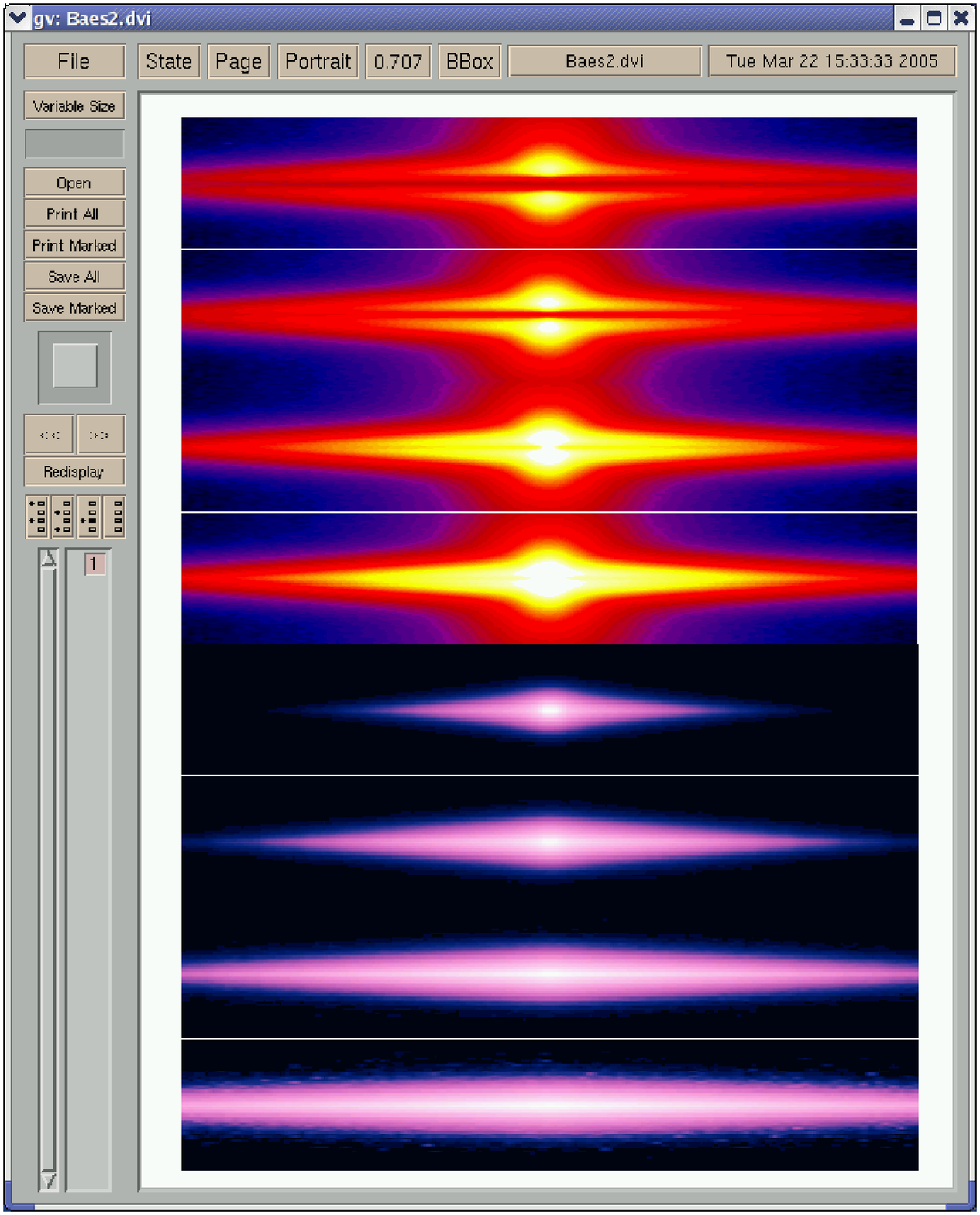}
\caption{Simulated maps of an edge-on spiral galaxy model similar to
NGC\,891 at various wavelengths ranging from the optical to the
submillimeter (from top to bottom: B, I, H, K, 60~$\mu$m, 100~$\mu$m,
200~$\mu$m and 850~$\mu$m). The maps are created by SKIRT from a model
including an double exponential disk and a de Vaucouleurs bulge with
typical stellar SEDs generated with the P\'{E}GASE stellar population
synthesis models. The dust is distributed in an exponential disk with
a scalelength $h_\txd=1.4\,h_*$, a scaleheight $z_\txd=0.6\,z_*$ and a
face-on optical depth $\tau_\txV=0.9$. The effects of absorption,
scattering and re-emission are properly taken into account.}
\label{f3.eps}
\end{figure}

During the last decade of the past century there has been a vivid
discussion about the opacity of spiral galaxies. This discussion was
initiated by Disney \cite{1989MNRAS.239..939D} and
Valentijn~\cite{1990Natur.346..153V}, who countered the conventional
view that spiral galaxies are optically thin over their entire optical
discs. The most reliable technique to model the dust content of spiral
galaxies is to combine imaging and spectra at wavelengths ranging from
the optical to the submillimeter. In particular, optical and
near-infrared data can be used to model the dust absorption and
scattering, whereas FIR/submm maps directly show the thermal emission
from the dust. Most of the detailed modelling efforts have
concentrated on nearby edge-on galaxies, where the appearance of a
clear dust lane in the optical images puts tight constraints on the
dust (e.g.\ \cite{1987ApJ...317..637K}, \cite{1997A&A...325..135X},
\cite{1998A&A...331..894X}, \cite{1999A&A...344..868X},
\cite{2000A&A...356..795A}, \cite{2000A&A...362..138P},
\cite{2001ApJ...548..150M}, \cite{2001A&A...372..775M},
\cite{2004A&A...414...45P}). The number of galaxies for which such
studies have been performed is still rather modest however. We are
planning to perform a multiwavelength study of a sample of edge-on
galaxies in order to investigate the dust distribution, optical depth,
stellar SEDs and star formation of spiral galaxies as a function of
Hubble type and surface brightness class. We hope to achieve this goal
by simultaneous modelling of the optical/NIR images and FIR/submm maps
or fluxes with the SKIRT code. An example of the output of the SKIRT
code is provided in figure~{\ref{f3.eps}}, which shows simulated maps
of an edge-on galaxy at optical, NIR, FIR and submillimeter
wavelengths. Our goal is to fit such maps to the observed data and
determine the model parameters using non-linear optimization
algorithms. Our first test results using genetic algorithms are
promising.

\section{Application~2: Elliptical galaxies kinematics}

During the past few years, a consensus has developed that elliptical
galaxies, like spiral galaxies, contain dark halos. Stellar kinematics
are generally considered as the most important tracer for these halos
at a few effective radii. Several authors have adopted stellar
kinematics to constrain the dark matter distribution in a number of
elliptical galaxies (\cite{1997ApJ...488..702R},
\cite{1998MNRAS.295..197G}). A possible caveat in these studies is the
effect of dust attenuation. Estimates based on IRAS data suggested
dust masses in ellipticals of the order of $10^5$ to $10^6$~M$_\odot$
\cite{1995A&A...298..784G}, but more recent ISO, SCUBA and NOBA data
increased typical dust mass estimates to about $10^7$~M$_\odot$
(\cite{2004ApJS..151..237T}, \cite{2004ApJ...612..837L},
\cite{2004dmu..conf...79V}).

\begin{figure}[t!]
\centering
\includegraphics[clip, bb=101 557 468 681,width=\textwidth]{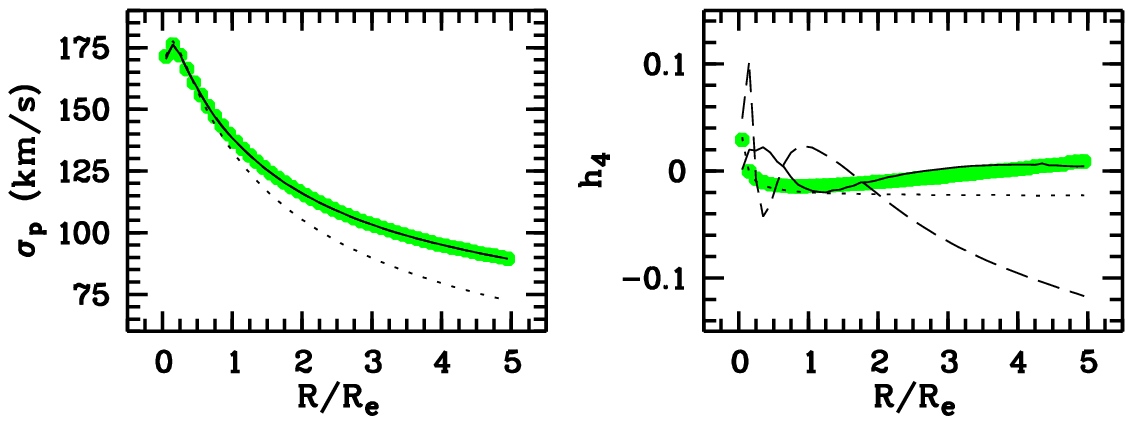}
\caption{The projected dispersion profile and the Gauss-Hermite $h_4$
profile for a simple dusty elliptical galaxy model with an optical
depth $\tau_\txV=1$. The dotted lines and the thick grey lines
represent the projected kinematics of the input model, respectively
without and with dust attenuation taken into account. The latter
results are used as input for the dynamical modelling procedure. The
dashed lines represent the best fitting model with constant $M/L$; the
solid line is the best fitting model with a dark matter halo. For more
details, see \cite{2001ApJ...563L..19B}.}
\label{f4.eps}
\end{figure}

We have done an extensive investigation of the effects of dust
attenuation on the observed kinematics of elliptical galaxies
(\cite{2000MNRAS.313..153B}, \cite{2000MNRAS.318..798B},
\cite{2001ApJ...563L..19B}, \cite{2002MNRAS.335..441B}). For a simple
elliptical galaxy model with a smooth dust component with optical
depth around unity, the effect of dust attenuation on the observed
kinematics is illustrated in figure~{\ref{f4.eps}}. For the lines of
sight that pass near the galaxy center, the kinematics are only
slightly affected. At large projected radii, however, the kinematics
are seriously affected: the velocity dispersion profile drops less
steeply, and the Gauss-Hermite $h_4$ parameter (a measure for the
shape of the line profiles) is significantly larger compared to the
dust-free case. These effects are caused by photons emitted by
high-velocity stars in the center of the galaxy, that are scattered in
the outskirts of the galaxy into lines of sight with a large projected
radii. Curiously, these effects are strikingly similar to the
kinematical signature of a dark halo, which is characterized by a
slowly decreasing dispersion profile and a positive $h_4$ profile.

To check this into more detail, we considered our dust-affected
kinematics as an observational data set, and modelled it in the usual
way, i.e.\ without taking dust attenuation into account. We found that
it was impossible to fit both the photometry and the kinematics with a
constant mass-to-light ratio model. A dark matter halo is hence
necessary to explain the effects caused by dust attenuation. In the
best fitting model with a dark halo, the dark matter contributes
roughly a third of the total mass within 1\,$R_e$, and half of the
total mass within the last data point. These results clearly
demonstrate that the effects of dust attenuation can mimic the
presence of a dark matter halo. In analogy with the mass-anisotropy
degeneracy, we are now faced with a new degeneracy, which could be
called the mass-dust degeneracy.  The new mass-dust degeneracy
strongly complicates the use of stellar kinematics as a tracer for the
mass distribution in elliptical galaxies: taking dust attenuation into
account in dynamical modelling procedures will reduce or may even
eliminate the need for a dark matter halo at a few $R_e$.

\section{Application~3: Spiral galaxy rotation curves}

\begin{figure}[t!]
\centering
\includegraphics[clip, bb=43 148 547 493, width=\textwidth]{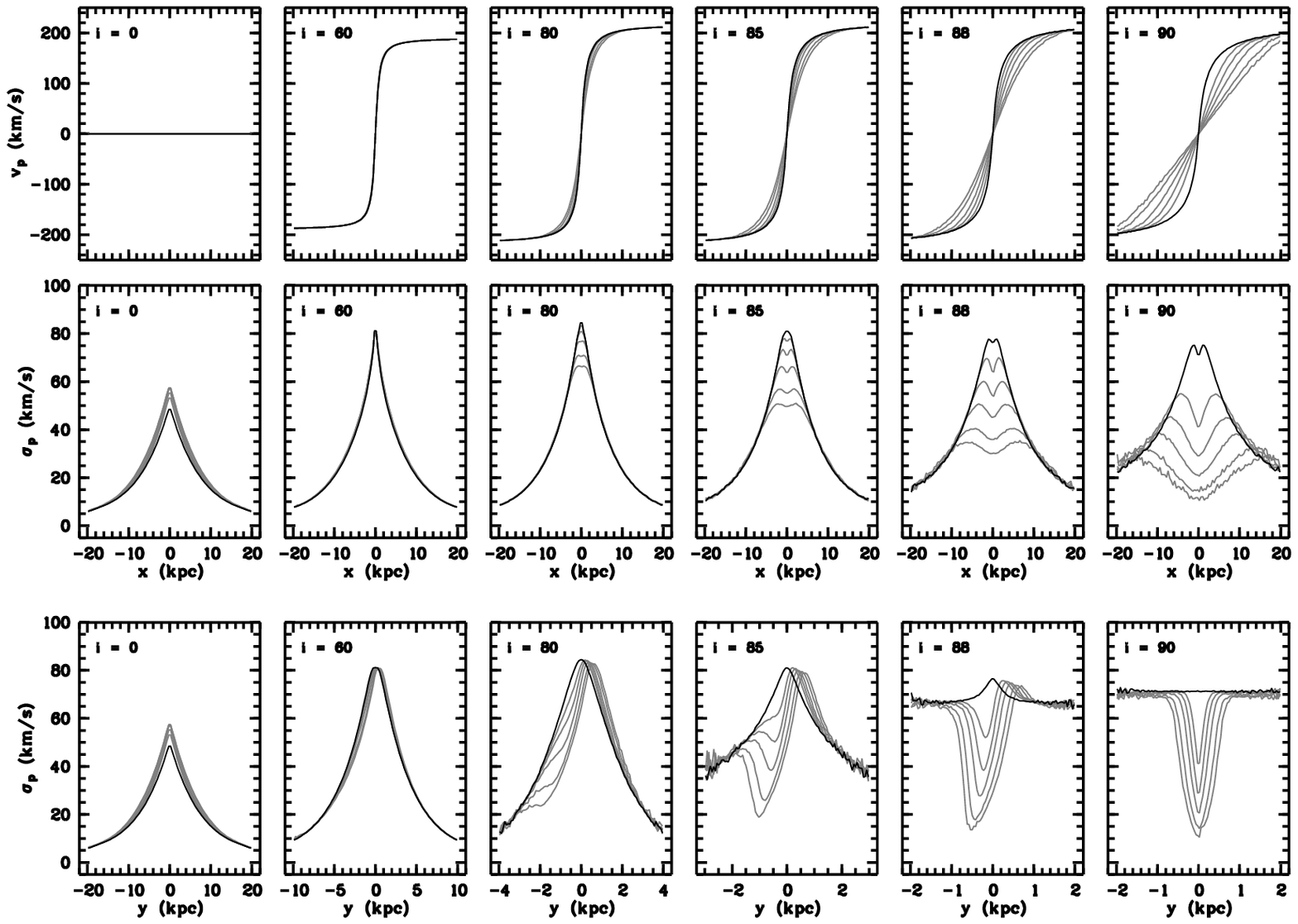}
\caption{The observed kinematics of our galaxy disc models along the
major and minor axes. The panels on the two upper rows represent the
mean projected velocity profiles and projected velocity dispersion
profiles along the major axis, and the bottom row show panels the
minor axis projected velocity dispersion profiles. The different
columns correspond to inclinations of 0, 60, 80, 85, 88 and 90 degrees
(from left to right). The different curves in each panel correspond to
different values of the optical depth: models with $\tau_\txV=0$
(black), 0.5, 1, 2, 5 and 10 are shown. For more details see
\cite{2003MNRAS.343.1081B}.}
\label{f5.eps}
\end{figure}

The observed flatness of spiral galaxy H{\sc i} rotation curves out to
very large distances clearly demonstrates the presence of dark
matter. Less clear, however, is the amount of dark matter present in
the inner regions of spiral galaxies. Various authors have reported a
discrepancy between the observed shallow slope of dark matter haloes
in LSB galaxies and the steep slope predicted by CDM cosmological
simulations (e.g.\ \cite{2003MNRAS.340..657D},
\cite{2004MNRAS.351..903G}). Important here is that the inner dark
matter profiles are usually derived from H$\alpha$ observations, which
are subject to dust attenuation.

We have performed radiative transfer simulations with SKIRT to
investigate the effect of dust attenuation on the observed (gas and
stellar) kinematics of spiral galaxies and to investigate whether dust
can serve as a valuable explanation for this discrepancy
\cite{2003MNRAS.343.1081B}. Figure~{\ref{f5.eps}} shows the major and
minor axis kinematics (mean projected velocity and velocity
dispersion) for a set of spiral galaxy models at various inclinations
and with various optical depths. For edge-on galaxies, there is a
strong effect of dust attenuation on the observed kinematics. The
dispersion profile shows a serious dip in the dust lane, such that
stellar kinematics measured through a dust lane should always be
treated with much caution. The rotation curve of edge-on spiral
galaxies becomes increasingly shallower for increasing optical depths,
an effect which is mainly caused by absorption (see also
\cite{1992ApJ...400L..21B}, \cite{2001ApJ...548..150M},
\cite{2003AstL...29..437Z}). Edge-on galaxies with an observed shallow
rotation curve could hence in reality have an intrinsically steep
cusp. However, this effect is strongly reduced when the galaxies are
more than a few degrees from exactly edge-on. For inclinations as
large as 80 degrees, dust attenuation is completely negligible. On the
one hand, this means that inclined galaxies form the ideal targets for
stellar kinematical studies, in particular to constrain the mass
structure and to study the kinematical history of disc galaxies. On
the other hand, it also means that dust attenuation cannot be invoked
as a possible mechanism to reconcile the discrepancies between the
observed shallow slopes of LSB galaxy rotation curves and the dark
matter cusps found in CDM cosmological simulations.

\end{document}